# Two-state protein-like folding of a homopolymer chain

Mark P. Taylor [a,*], Wolfgang Paul [b,c], and Kurt Binder [b]

[a]*Department of Physics, Hiram College, Hiram, OH 44234, USA*
[b]*Institut für Physik, Johannes-Gutenberg-Universität, Staudinger Weg 7, D-55099 Mainz, Germany*
[c]*Institut für Physik, Martin-Luther-Universität, D-06099 Halle (Saale), Germany*

Many small proteins fold via a first-order "all-or-none" transition directly from an expanded coil to a compact native state. Here we study an analogous direct freezing transition from an expanded coil to a compact crystallite for a simple flexible homopolymer. Wang-Landau sampling is used to construct the 1D density of states for square-well chains of length 128. Analysis within both the micro-canonical and canonical ensembles shows that, for a chain with sufficiently short-range interactions, the usual polymer collapse transition is preempted by a direct freezing or "folding" transition. A 2D free-energy landscape, built via subsequent multi-canonical sampling, reveals a dominant folding pathway over a single free-energy barrier. This barrier separates a high entropy ensemble of unfolded states from a low entropy set of crystallite states and the transition proceeds via the formation of a transition-state folding nucleus. Despite the non-unique homopolymer ground state, the thermodynamics of this direct freezing transition are identical to the thermodynamics of two-state protein folding. The model chain satisfies the van't Hoff calorimetric criterion for two-state folding and an Arrhenius analysis of the folding/unfolding free energy barrier yields a Chevron plot characteristic of small proteins.

## 1. Introduction

Macromolecules undergo conformational transitions that may be viewed as small system analogs to the phase transitions exhibited by a bulk fluid. Thus the polymer collapse or coil-globule transition can be compared to the gas-liquid (condensation) transition where, in each of these cases, we go from a disordered expanded (low-density) state to a disordered compact (condensed) state. Similarly, the crystallization of a single macromolecule is a small system analog to the freezing transition of a bulk system. Since macromolecules are inherently finite size systems some caution must be exercised when comparing single-molecule conformational transitions with bulk system phase transitions. In the case of synthetic polymers one can appeal to a hypothetical thermodynamic or infinite chain length limit and in this limit one finds that the polymer coil-globule transition becomes a true second order phase transition similar to a gas-liquid critical point [1]. In the case of biological macromolecules such a proteins, which are heteropolymers of fixed length, no such thermodynamic limit exits, even hypothetically. However, in such cases one can still make use of a microcanonical analysis to define phase transitions in finite size systems [2,3].

* Corresponding author. Tel.: 330-324-5241; fax: 330-342-5448; *Email address*: taylormp@hiram.edu.

A connection between the conformational transitions of a homopolymer chain and the protein folding transition was suggested by Zhou, Hall, and Karplus more than a dozen years ago [4,5]. These authors carried out computer simulations of a flexible homopolymer chain that underwent a collapse transition followed by a low temperature freezing transition. They proposed that the collapsed homopolymer provided a model for the molten globule state of a protein and that the freezing transition was analogous to a molten globule to native state folding transition. The recent work of Rampf, Paul, and Binder has renewed interest in single chain phase transitions for simple homopolymers [6,7]. In particular, these authors suggested that a direct transition from an expanded coil to a frozen crystallite might be possible in a system with sufficiently short-range interaction. Taylor, Paul, and Binder have recently confirmed this prediction and have suggested that such a direct freezing transition may provide a simple model for the "all-or-none", or expanded coil to native state, folding transition found for many small proteins [8,9].

Here we review the recent findings of Taylor, Paul, and Binder and carry out further analysis of the direct freezing or homopolymer folding transition. In particular we construct a two-dimensional free energy landscape for this folding process and compare the thermodynamics of this model with results from a simple two-state analysis as is commonly used in the analysis of protein folding kinetics and thermodynamics.

**2. Theory**

*2.1. Chain Model*

In this work we study a single linear homopolymer chain at temperature $T$. The chain is comprised of $N$ spherically symmetric interaction-site monomers connected by completely flexible "universal joints" of bond length $L$. Non-bonded monomers interact through the square-well (SW) potential

$$u(r) = \begin{cases} \infty & r < \sigma \\ -\varepsilon & \sigma < r < \lambda\sigma \\ 0 & r > \lambda\sigma \end{cases} \tag{1}$$

where $r$ is the monomer-monomer separation, $\sigma$ and $\lambda\sigma$ are the hard sphere and square well diameters, respectively, and $\varepsilon$ is the well depth. We define a reduced temperature $T^* = k_B T/\varepsilon$, where $k_B$ is the Boltzmann constant, and take the bond length to be $L = \sigma$. The square-well chain model has a discrete energy spectrum $E = -n\varepsilon$ where $n$ is the number of square-well overlaps in a chain conformation, such that $0 \leq n \leq n_{max}$, and the chain ground-state energy (which depends on $\lambda$) is $E_{gs} = -n_{max}\varepsilon$. The probability to find the chain in a particular energy state $E$ is given by

$$P(E,T) = \frac{1}{Z(T)} g(E) e^{-E/k_B T} \tag{2}$$

where $Z(T)$ is the canonical partition function given by

$$Z(T) = \sum_n g(E_n) e^{-E_n/k_B T}. \tag{3}$$

and $g(E)$ is the energy density of states. As suggested by Eqs. (2) and (3), $g(E)$ provides access to the complete thermodynamics of the SW-chain. This density of states function is formally defined by a $3N$-dimensional integral over all chain conformations [10] and gives the volume of configuration space associated with energy state $E$. While direct calculation of $g(E)$ is impractical for all but the shortest chains ($N \leq 6$) [10,11] this function can be obtained for longer chains through advanced simulation techniques.

*2.2. Simulation Methods*

Here we use the Wang-Landau (WL) algorithm [12] to construct density of states functions for single SW chains. In this approach one generates a sequence of chain conformations using a set of Monte Carlo (MC) moves; however, rather than accepting moves according to a temperature dependent Boltzmann weight (i.e., via the Metropolis criterion), one uses the following temperature independent multicanonical acceptance probability:

$$P_{acc}(a \to b) = \min\left(1, \frac{w_b g(E_a)}{w_a g(E_b)}\right) \quad (4)$$

where $w_a$ and $w_b$ are conformation dependent weight factors which ensure microscopic reversibility for a given MC move [13-15]. If $g(E)$ were known this approach would yield a uniform exploration of all energy states of the system (and thus, presumably of all configurational phase space). In the WL approach $g(E)$ is constructed in an iterative and dynamic fashion where smaller scale refinements are made at each level of iteration. Thus, in the $m$th level of the iteration, after every attempted MC move the density of states associated with the current chain energy is modified by a multiplicative factor $f_m>1$ via $g(E_n) \to f_m g(E_n)$ and an energy state visitation histogram is simultaneously incremented via $H(E_n) \to H(E_n)+1$. This $H(E_n)$ histogram is periodically checked for "flatness" [*i.e.*, each entry in $H(E_n)$ is within $p$ percent of the overall average value of $H(E_n)$] indicating an approximate equal visitation of all energy states. When flatness is achieved the modification factor is reduced via $f_{m+1} = \sqrt{f_m}$, the visitation histogram is reset to zero for all states and the $(m+1)$st level of iteration begins. We also employ an alternate "uniform growth" flatness criteria which requires that each entry in $H(E_n)$ has increased by an amount within $p$ percent of the overall average increase in $H(E_n)$. This avoids convergence difficulties associated with the generation of highly asymmetric histograms that can occur due to "bottlenecks" in configuration space [8,9].

In our implementation of the WL algorithm we begin with an initial guess of $g(E_n) = 1$ for all states, take $f_0 = e^1$, $p = 20$, and check for flatness and uniform growth every $10^4$ and $5 \times 10^7$ MC cycles, respectively. For our continuum chain model we find that at least 26 levels of iteration ($f_{26}-1 \approx 10^{-8}$) are required to achieve a converged solution and in most cases we continue the simulation to $m = 30$ ($f_{30}-1 \approx 10^{-9}$). In the WL algorithm the values for $g(E_n)$ continuously grow through the simulation as the density of states function is determined only up to a multiplicative constant. To avoid numerical difficulties we actually compute the logarithm of $g(E_n)$ and at every flatness check we uniformly reduce all entries in the $\ln[g(E_n)]$ estimate by the current maximum entry.

Successful application of the WL approach requires an MC move set that is capable of efficiently exploring all configuration space. We use a combination of single interior-bead crankshaft and end-bead rotation and reptation moves along with multi-bead pivot and end-bridging moves. For the end-bridging move we randomly select end site 1 (or $N$), identify all interior sites $i > 3$ (or $i < N-2$) within distance $2L$ of this end site, and selecting one of these at random, join sites $i$ and 1 (or $N$) via removal and reinsertion (at a randomly chosen azimuthal angle) of site $i-1$ (or $i+1$). The weight factor for this bond-bridging move is $w_b = b_a J_b$ where $b_a$ is the number of possible bridging sites $i$ present in state $a$ and $J_b = 1/r_{1i}$ (or $1/r_{iN}$) is a Jacobian factor arising from the fixed bond length restriction [13,15]. Weight factors for the other MC moves are all unity. A single MC cycle involves all of these move types [9].

We have verified the WL algorithm with this MC move set for the SW chain model by direct comparison with exact $g(E_n)$ results for short ($N \leq 6$) chains [10] for $1.01 \leq \lambda \leq 1.9$ and by comparison with the Metropolis MC results of Zhou *et al.* [5] for SW chains with $N = 64$ and $\lambda = 1.5$. For longer chains we find it most efficient to carry out a set of WL simulations for a set of overlapping energy windows [6]. MC moves taking the system out of the window are rejected with no updating of $g(E_n)$ or $H(E_n)$ (note however, that moves resulting in hardcore overlap are always rejected *with* updating). Results from these overlapping regions are joined by matching $g(E_n)$ values at the center of the overlap region. Before we undertake a full WL simulation we carry out a preliminary run at low energies, without a low energy cutoff, to estimate the ground state energy of the chain. The $H(E)$ histogram generated in this preliminary run allows us to estimate the lowest energy state we can include in our full simulation which will still allow the WL method to converge in a "reasonable" amount of time. For longer chains our lowest energy window typically extends to within a few percent of our estimated ground state energy [9].

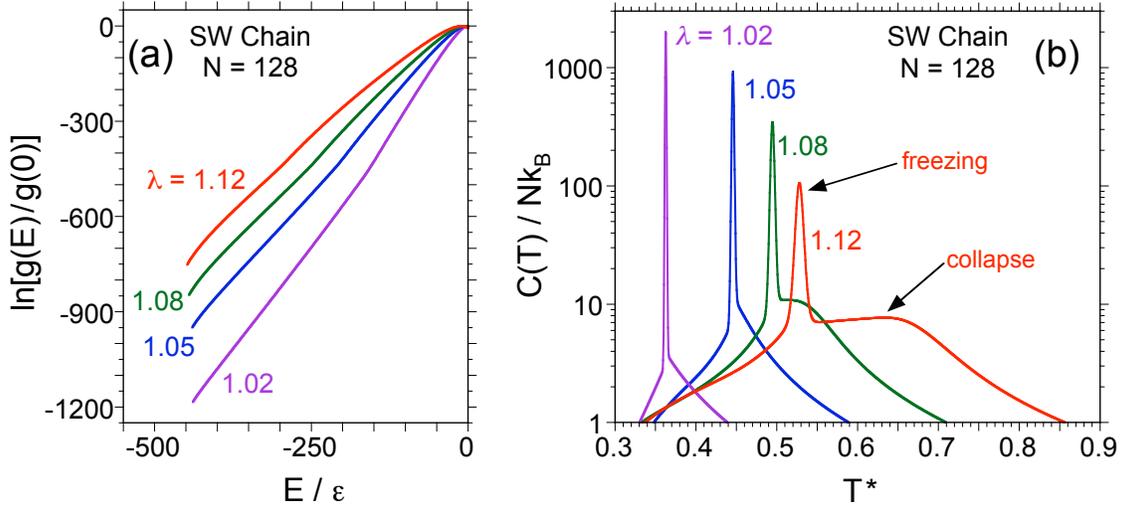

FIG. 1 (a) Logarithm of the density of states $\ln[g(E)]$, relative to the value at $E=0$, versus reduced energy $E/\varepsilon$ and (b) canonical specific heat per monomer $C(T)/Nk_B$ versus reduced temperature $T^*$ for a SW chain of length $N = 128$ and well diameters $\lambda\sigma$ as indicated.

## 3. Results

### 3.1. Density of States

In Fig. 1(a) we show density of states results obtained via Wang-Landau sampling for SW chains of length $N = 128$ for a range of well diameters $1.02 \leq \lambda \leq 1.12$. Results are given in logarithmic form relative to the value at energy $E = 0$. Each of these results is an average over two to four independent sets of simulations where each set consists of two overlapping energy windows $E \in [0, E_{join}–50\varepsilon]$ and $[E_{join}+50\varepsilon, E_{min}]$ where $E_{join}$ is the energy at which results from the two windows are joined. For these results $E_{join}$ is near $-250\varepsilon$ and $E_{min}$ ranges from $-438\varepsilon$ to $-448\varepsilon$ (being approximately $10\varepsilon$ above our best estimate of the ground state energy in each case). We assess the quality of these joins by examining the derivative $d\ln g(E)/dE$ in the neighborhood of $E_{join}$ and find very good agreement between the two windows in all cases. The resulting $g(E)$ functions are quite reproducible between independent runs, except in a narrow energy range associated with the formation of partially crystalline transition state [8,9]. Despite this small region of discrepancy, all transition temperatures obtained from these functions agree to within $\delta T^* = \pm 0.006$ between independent runs. These density of states functions span hundreds of orders of magnitude (approximately 500 in the case of $\lambda = 1.02$) encompassing the full range of chain conformations, from the highly expanded coil to the compact crystallite.

### 3.2. Canonical Analysis

From the single chain density of states function $g(E)$ we construct the canonical partition function $Z(T)$ and canonical probability function $P(E,T)$ as given in Eqs. (3) and (2), respectively. From these we obtain the chain internal energy

$$U(T) = \langle E \rangle = \sum_n E_n P(E_n, T). \qquad (5)$$

Although the WL simulation method only gives $g(E)$, and thus $Z(T)$, up to an arbitrary multiplicative constant, the resulting canonical probability function $P(E,T)$ (and averages over functions of $E$ such as $U(T)$) are determined absolutely. To locate phase transitions within the canonical ensemble one typically examines the specific heat function

$$C(T) = \frac{dU}{dT} = \frac{1}{k_B T^2}\left(\langle E^2 \rangle - \langle E \rangle^2\right). \tag{6}$$

For finite size systems, maxima in $C(T)$ can be associated with phase transitions or other structural rearrangements [4,5,14]. Discontinuous (first-order) and continuous transitions are distinguished by a bimodal versus unimodal $P(E,T)$ distribution, respectively, and coexisting states for a discontinuous transition are identified via the condition of equally weighted peaks in the bimodal $P(E,T)$ function.

In Fig. 1(b) we show canonical specific heat functions computed from the density of states functions shown in Fig. 1(a). Each of these functions displays a strong low temperature peak that we identify with a chain freezing transition. The first-order character of this transition is confirmed by bimodal $P(E,T)$ distributions for the temperatures corresponding to these $C(T)$ peaks. For the cases of $\lambda = 1.12$ and $1.08$ there is also a high temperature peak in the $C(T)$ functions. This rather broad peak can be identified with a chain collapse transition and the corresponding unimodal $P(E,T)$ function verifies the continuous nature of this transition. No distinct high temperature collapse transition peak is observed in the $C(T)$ functions for $\lambda = 1.05$ and $1.02$. This suggests that there may be a "direct" freezing transition here from the expanded coil state, or it could be that the collapse peak is simply "hidden" under the very strong freezing peak. Unfortunately, these two possibilities cannot be distinguished within a canonical ensemble analysis.

*3.3. Microcanonical Analysis*

To clarify the issue of a direct freezing transition versus a "hidden" collapse transition we turn to a thermodynamic analysis within the microcanonical ensemble [2,3,16,17]. In this approach we examine the curvature properties of the microcanonical entropy function given by

$$S(E) = k_B \ln g(E). \tag{7}$$

The first derivative of $S(E)$ gives the microcanonical temperature via

$$T(E) = [dS/dE]^{-1} \tag{8}$$

while the second derivative of $S(E)$ gives the microcanonical specific heat

$$c(E) = [dT/dE]^{-1} = \frac{-1}{T^2(E)}\left(\frac{d^2S}{dE^2}\right)^{-1}. \tag{9}$$

A discontinuous transition in the microcanonical ensemble is signaled by a convex intruder in the entropy function which appears as a Maxwell-like loop in the temperature function. Coexisting states can be identified through a double tangent construction on $S(E)$ or through an equivalent equal area construction on the inverse temperature function $T^{-1}(E)$. A continuous phase transition is signaled by an inflection point in the temperature function $T(E)$ which produces an isolated peak in the $c(E)$ specific heat function.

In Figs. 2(a) and 2(b) we show the inverse temperature functions for the SW $N = 128$ chains with well diameters $\lambda = 1.05$ and $1.08$, respectively. Both of these cases display a pronounced "loop" indicative of a discontinuous transition and the equal areas construction yields a transition temperature in agreement with the canonical ensemble $C(T)$ peak location. The corresponding microcanonical specific heat functions $c(E)$ are shown in Figs. 2(c) and 2(d). Both of these functions possess a set of poles, which are the signature for a discontinuous phase transition, coincident with the local maximum and minimum in the inverse temperature function. The $c(E)$ curves also display weak isolated peaks at energies above the freezing transition which can be associated with the continuous collapse transition. In the case of $\lambda = 1.08$ the collapse peak falls outside the two-phase coexistence region of the freezing transition and thus distinct collapse and freezing transitions are observed (as seen in the canonical specific heat function). However, in the case of $\lambda = 1.05$ the collapse peak is located within the two-phase coexistence region which means that the collapse transition will be preempted by the freezing transition. Thus, the SW chain with a

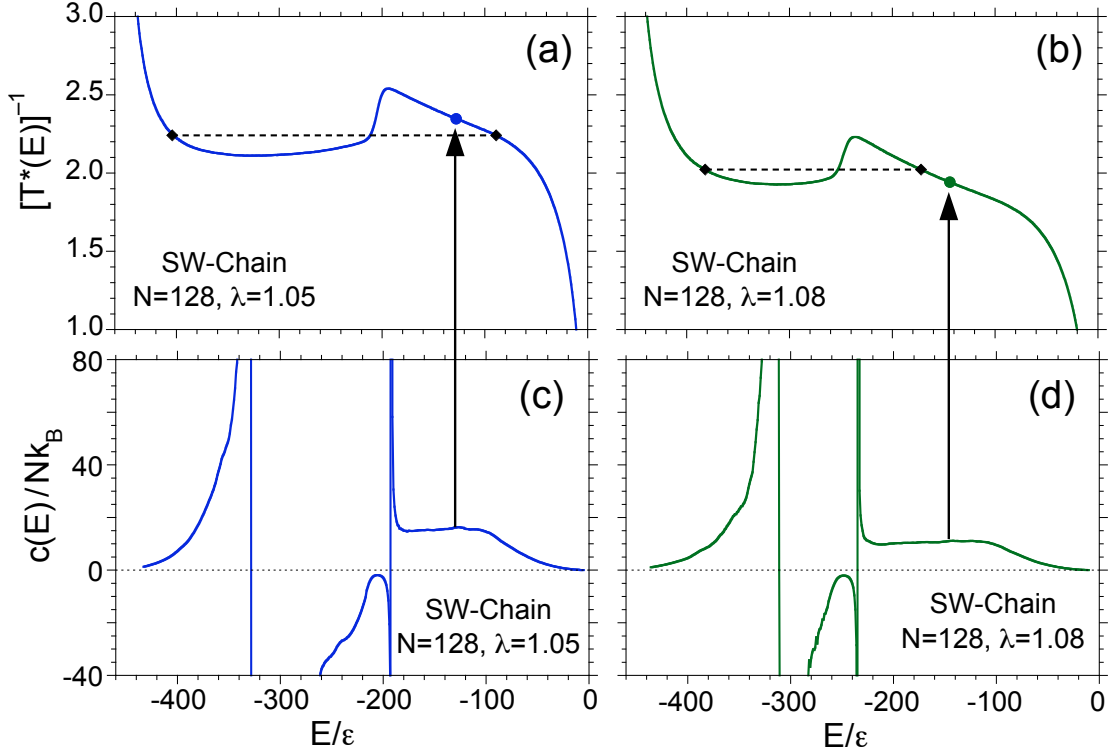

FIG. 2 (a,b) Microcanonical inverse temperature $T^{-1}(E)$ and (c,d) specific heat $c(E)/Nk_B$ versus reduced energy $E/\varepsilon$ for a SW chain of length $N = 128$ and well diameters $\lambda = 1.05$ and 1.08, as indicated. The "loops" in the inverse temperature curves signal a discontinuous phase transition and the dashed horizontal lines, obtained via an equal area construction, locate the coexisting phases. The weak maxima in the $c(E)$ functions locate the continuous collapse transition, shown as the filled circles on the inverse temperature curves. In the case of $\lambda = 1.05$ this transition is located within the two-phase coexistence region and thus chain collapse is preempted by a direct freezing transition from the expanded coil state.

sufficiently short-range interaction ($\lambda \leq 1.06$ for $N = 128$) displays a direct freezing transition from an expanded coil to a compact crystallite [8,9]. This result is analogous to the disappearance of a liquid like phase in the phase diagram of a simple fluid simple with very short-range interactions [18-20].

*3.4. Free Energy Landscape*

The direct freezing transition we have identified for flexible SW chains with short-range interactions is in many ways analogous to the "all-or-none" folding transition exhibited by many small proteins [21-23]. To explore this analogy we have constructed a free energy landscape of the homopolymer direct freezing or folding transition. To build this two-dimensional landscape we carry out a multicanonical simulation [24] using our WL $g(E)$ results as the multicanonical weights. In this simulation we construct a conditional probability function $P(Q|E_n)$ on the conformation dependent variable $Q$. From this temperature independent probability function on $Q$ one can build a two-dimensional free-energy surface via

$$G(E,Q,T) = -k_B T \ln[P(Q|E)P(E|T)]. \tag{10}$$

where $P(E|T)$ is the 1D canonical energy probability function given by Eq. (2).

We have chosen to build such a surface using chain size, measured by the radius of gyration squared $R_g^2$, as the second variable $Q$. We show part of this two-dimensional free energy surface in Fig. 3 for the $N = 128$ SW chain with $\lambda = 1.05$ at the direct freezing, or folding, transition temperature $T^* = 0.446$. This surface exhibits two global minima separated by a relatively simple barrier. The broad free energy basin in the vicinity of $110 > -E/\varepsilon > 75$ and $12 < R_g^2/\sigma^2 < 25$ can be associated with an ensemble of expanded coil states while the narrow deep well near

– 6 –

$E/\varepsilon \approx -400$, $R_g^2/\sigma^2 \approx 5$ corresponds the coexisting set of compact crystallite states. We have sketched the dominant folding pathway on this diagram, defined as the path of minimum local free energy connecting the two global minima. This path passes over the free energy barrier in a slightly indirect manner corresponding to the required structural transformation associated with the formation of a partially crystalline folding nucleus or transition state.

In Fig. 4(a) we show the projection of our 2D free energy landscape onto the $E$-axis. This one-dimensional free energy function is simply given $G(E,T) = E - TS(E)$. To facilitate comparison with biological systems, we map the reduced folding temperature $T^* = 0.446$ to a physical temperature of $T_{equil} = 60°C$ (333 K) which sets the SW energy parameter to $\varepsilon = 1.5$ kcal/mol. At this equilibrium temperature the free energy barrier separating the folded and unfolded states is approximately $\Delta G^\ddagger \approx 8\varepsilon \approx 12$ kcal/mol. Fig. 4(a) also shows the free energy function $G(E,T)$ at physically meaningful temperatures both above and below the equilibrium folding temperature. For $T < T_{equil}$ the folded state is stabilized and the free energy barrier to folding is lowered. Analogous behavior is exhibited for $T > T_{equil}$ where the folded state is destabilized. In Fig. 4(b) we show the canonical probability distribution $P(E,T)$ at the equilibrium folding temperature in which case the peaks of the bimodal distribution have equal areas.

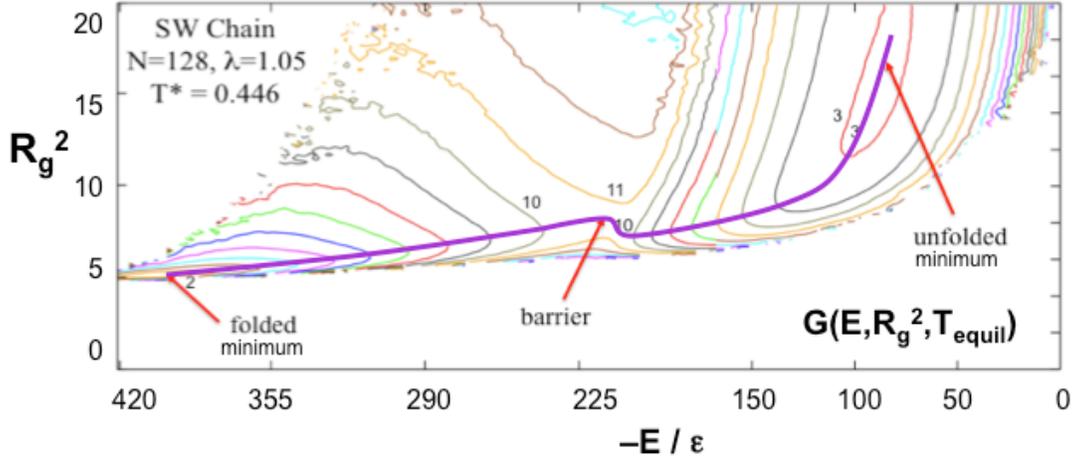

FIG. 3 Two-dimensional free energy landscape for a SW chain with length $N = 128$ and well diameter $\lambda = 1.05$ at the equilibrium folding temperature $T^* = 0.446$. Free energy contours are plotted as a function of chain radius of gyration squared $R_g^2$ and interaction energy $E$. The contour spacing is $1\varepsilon$. The solid line connecting the basin of unfolded states with the folded minimum represents the dominant folding pathway through this landscape.

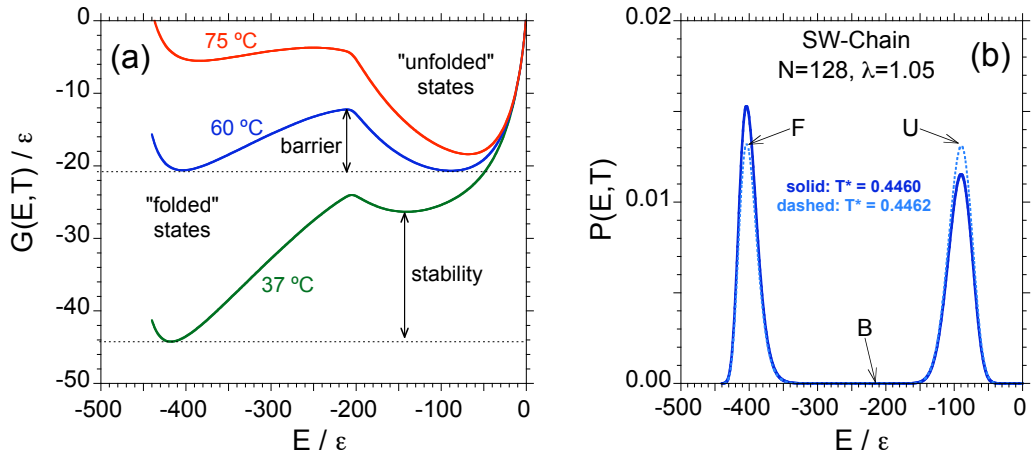

FIG. 4 (a) Free energy $G(E,T)/\varepsilon$, relative to value at $E=0$, and (b) canonical probability distribution $P(E,T)$ for a SW chain with length $N = 128$ and well diameter $\lambda = 1.05$ at temperatures as indicated. The equilibrium folding temperature is $T^* = 0.4460$ ($T = 60°C$) which corresponds to equal weighted $P(E,T)$ peaks. In (b) the dashed $P(E,T)$ curve corresponds to the equal height two-state approximation and the two-state folded (F), unfolded (U), and transition state or barrier (B) species are indicated.

*3.5. Two-state Approximation*

For the equilibrium free energy surface shown in Fig. 3 a high entropy ensemble of unfolded states is in equilibrium with a low entropy set of "folded" (i.e., crystallite) states. These sets of states are connected by a single dominant path in the landscape that passes over a single free-energy barrier. This behavior is completely analogous to the folding behavior exhibited by many small proteins which fold directly from an expanded coil to the compact (and unique) native state [21-23]. The thermodynamics and kinetics of such "all-or-none" protein folding is often described in terms of a simple two-state model in which one treats the ensemble of unfolded chain states as a unique species [23,25,26]. In this approximation we have the equilibrium F ⇔ U where the folded (F) and unfolded (U) states are each characterized by an enthalpy ($H_F$ or $H_U$) and entropy ($S_F$ or $S_U$). The equilibrium folding transition corresponds to the free energy equality $G_F = G_U$ where $G = H - TS$. The F and U states are separated by a single free energy barrier corresponding to a transition or barrier state B characterized by enthalpy $H_B$ and entropy $S_B$.

Here we explicitly test the accuracy of this two-state approximation for the direct freezing or homopolymer folding transition exhibited by the $N = 128$ SW chain with $\lambda = 1.05$. We take the U and F states to be defined by the two maxima in the bimodal probability function $P(E,T)$ for the condition that these two peaks have equal height (as opposed to equal area). As shown in Fig. 4(b), this two-state equal height condition gives an equilibrium transition temperature of $T^* = 0.4462$ (as opposed to the correct equal-areas transition temperature of $T^* = 0.4460$) with folded and unfolded species defined by $H_F = -404\varepsilon$, $S_F = -856.52 k_B$ and $H_U = -90\varepsilon$, $S_U = -152.74 k_B$, respectively, where we assume $H \approx E$ and the chain entropies are measured relative to the entropy of the $E = 0$ state. The barrier state B is defined by the $P(E,T)$ minimum separating the F and U states and is characterized by $H_B = -212\varepsilon$, $S_B = -444.86 k_B$.

The two-state version of the canonical partition function [Eq. (3)] is simply given by

$$Z_{2\text{state}}(T) = g(E_F)e^{-E_F/k_B T} + g(E_U)e^{-E_U/k_B T} = e^{-G_F/k_B T} + e^{-G_U/k_B T} \qquad (11)$$

which leads to the two-state version of the canonical specific heat [Eq. (6)]

$$C_{2\text{state}}(T) = \frac{1}{[Z_{2\text{state}}(T)]^2} \frac{(H_U - H_F)^2}{k_B T^2} e^{-(G_U + G_F)/k_B T}. \qquad (12)$$

In Fig. 5(a) we compare the canonical specific heat computed within this two-state approximation with the results obtained from the full $g(E)$ (shown in Fig. 1(b)). The two-state approximation clearly provides a very good representation of the specific heat function for the homopolymer folding transition. Experimentally this two-state approximation is often assessed through a van't Hoff analysis [23] in which the enthalpy difference between the two states is written as

$$\Delta H_{vH} = -k_B T_m^2 (d \ln K / dT)_{T=T_m} = -2 T_m [k_B C(T_m)]^{1/2} \qquad (13)$$

where $K = [U]/[F] = \exp\{-(G_U - G_F)/k_B T\}$ is the equilibrium constant for the F ⇔ U "reaction" and $T_m$ is the temperature of the transition midpoint where $[U] = [F]$ (and thus $G_F = G_U$). In practice, $T_m$ is usually taken as the peak location of the specific heat $C(T)$. The equality of the van't Hoff enthalpy $\Delta H_{vH}$ given by Eq. (13) and the calorimetric enthalpy $\Delta H_{cal}$, given by the area under the $C(T)$ curve, is taken as evidence for two-state folding. (This condition is satisfied exactly by the pure two-state model as can be seen by inserting Eq. (12) into Eq. (13)). In the case of the full SW chain model with $N = 128$ and $\lambda = 1.05$ we find $\Delta H_{cal} = 306\varepsilon$ and $\Delta H_{vH} = 307\varepsilon$ which would be considered to satisfy the experimental calorimetric criterion of $\Delta H_{vH} / \Delta H_{cal} \approx 1.0$.

An alternate experimental test of two-state folding behavior is to study the temperature dependence of the folding and unfolding kinetics [23,26]. In the two-state approximation the kinetics of these processes is assumed to follow a simple Arrhenius law where the folding (or unfolding) rate is given by

$$k_{rate} = a e^{-\Delta G^\ddagger / k_B T} \qquad (14)$$

where $\Delta G^\ddagger = \Delta H^\ddagger - T\Delta S^\ddagger$ is the height of the free energy barrier from the initial state (i.e., $\Delta G_{\text{fold}}^\ddagger \equiv G_B - G_U$) and $a$ is the elementary rate of barrier crossing events. In the two-state model a plot of $-\Delta G^\ddagger / k_B T$ versus $1/T$ will yield straight lines for folding and unfolding with slopes $-\Delta H_U^\ddagger = H_U - H_B$ and $-\Delta H_F^\ddagger = H_F - H_B$, respectively. Experimentally, such so called Chevron plots are constructed by plotting the logarithm of the measured folding and unfolding rates $\ln(k_{rate})$ versus $1/T$ or versus denaturant concentration. In either case, the linearity of the folding and unfolding branches is taken as evidence for two-state folding [23,26]. In Fig. 5(b) we show a Chevron plot for the SW chain model with $N = 128$ and $\lambda = 1.05$ computed both within the two-state approximation and for the full model. In the two-state approximation the folding and unfolding branches are simply given by $-\Delta G_{unfold}^{\ddagger}/T^* = -\Delta H_F^{\ddagger}/T^* + \Delta S_F^{\ddagger}/k_B$ and $-\Delta G_{fold}^{\ddagger}/T^* = -\Delta H_U^{\ddagger}/T^* + \Delta S_U^{\ddagger}/k_B$ respectively, where the required enthalpies and entropies defining the F, U, and B species are given above. The crossing of the folding and unfolding branches in the Chevron plot corresponds to equilibrium conditions. As seen in Fig. 5(b), the two-state approximation provides a good representation of the kinetics of the full model in the vicinity of this equilibrium point. However, away from the equilibrium point the full model displays deviations from linearity, especially on the folding branch. Such non-linearity or Chevron rollover is frequently encountered in experimental Chevron plots, especially on the folding side, and is often taken as evidence of non-two-state folding due to the appearance of a stable intermediate on the folding pathway [26,27]. In our model homopolymer system the Chevron rollover on folding is due primarily to the temperature evolution of the unfolded states (U), which move to an average lower energy with decreasing temperature (*i.e.*, $|\Delta H_U^{\ddagger}|$ decreases for $T < T_{equil}$ as seen in Fig. 4(a)) due to chain size reduction in a worsening solvent. Given the commonality of the homopolymer and protein unfolded states, our results suggest that a Chevron rollover on folding is to be expected and thus does not necessarily indicate a break down of the two-state-like process. The homopolymer results shown in Fig. 5(b) also show a mild Chevron rollover on the unfolding side. This is primarily due to movement of the transition state (B) to lower energies with increasing temperature (*i.e.*, $|\Delta H_F^{\ddagger}|$ decreases for $T > T_{equil}$ as seen in Fig. 4(a)). Such transition state movement and broadening has also been suggested to explain Chevron rollovers in experimental systems with chemical denaturants [28].

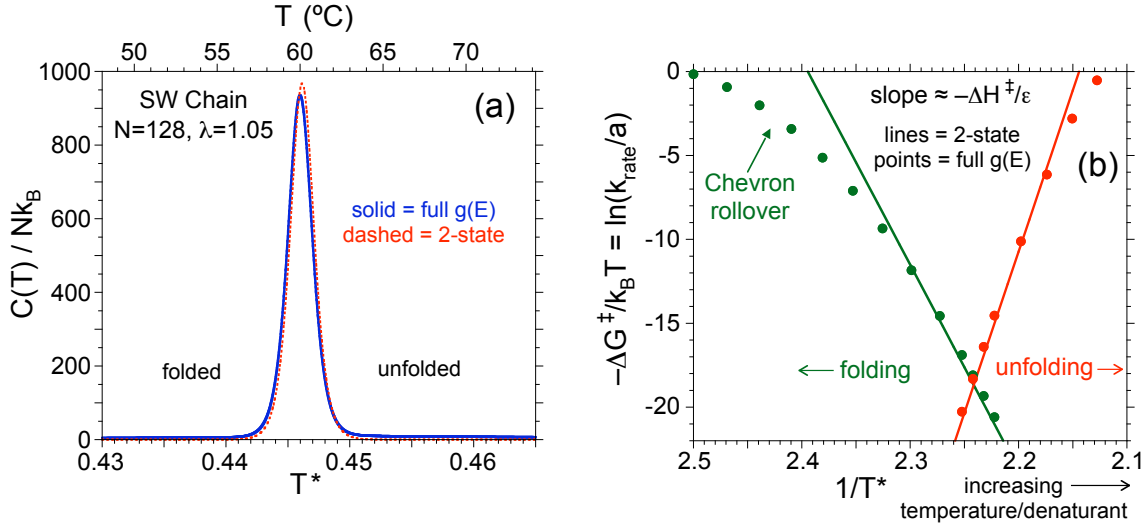

FIG. 5 (a) Canonical specific heat per monomer $C(T)/Nk_BT$ versus reduced temperature $T^*$ and (b) Chevron plot showing the folding and unfolding free energy barriers $-\Delta G^{\ddagger}/k_BT$ versus inverse reduced temperature $1/T^*$ for a SW chain of length $N = 128$ and interaction range $\lambda = 1.05$. In (a) the dashed line shows the two-state approximation result $C_{2state}(T)$ [Eq. (12)] while the solid line shows the $C(T)$ computed using the complete density of states (shown in Fig. 1). The physical temperature scale shown on the upper axis assumes an equilibrium folding temperature of 60 ºC. In (b) the straight lines show results from the simple two-state approximation while the symbols give results computed using the complete density of states.

## 4. Conclusions

The conformational phase behavior of a flexible homopolymer chain displays a number of interesting features. Chains with a not too short interaction range undergo the transition sequence on cooling of expanded coil→collapsed globule→frozen crystallite [5,9,29]. As we have recently shown and further explore in this work, for chains with sufficiently short range interaction the chain collapse transition disappears from this sequence being preempted by a direct discontinuous expanded coil→frozen crystallite transition. This direct freezing or homopolymer folding transition is analogous to the "all-or-none" folding transition exhibited by many small proteins. As we have described here, the simple homopolymer model reproduces the key thermodynamic features that provide the experimental signatures for two-state protein folding. Thus, neither a unique ground state nor complex interactions are required for a chain model to display two-state folding. Our study shows that a sufficiently large free energy barrier separating the ensembles of unfolded and folded states as a well as a sufficiently low entropy for the folded states is enough to produce protein folding thermodynamics. In our homopolymer model, a low-entropy folded state is produced by the short-range interaction while a large free energy barrier to folding is created by the required formation of a partially crystalline transition state structure on the folding pathway. This transition state structure is consistent with the nucleation event associated with a first-order or discontinuous phase transition and thus the homopolymer freezes (folds) via a nucleation-condensation mechanism of the type proposed for fast folding proteins [30].

## Acknowledgements

The authors thank Jutta Luettmer-Strathmann, Michael Bachmann, Peter Virnau, and David Landau for helpful discussions. Financial support through the Deutsche Forschungsgemeinschaft (SFB 625/A3) and the National Science Foundation (DMR-0804370), and a sabbatical leave from Hiram College are also gratefully acknowledged.